# Granular Effect on Electron Conduction in Discontinuous Metal Films

Juhn-Jong Lin[1 a)], Zhi-Qing Li[2 b)], Chih-Yuan Wu[3] and Sheng-Shiuan Yeh[4]

[1]Department of Electrophysics, National Yang Ming Chiao Tung University, Hsinchu 30010, Taiwan
[2]Tianjin Key Laboratory of Low Dimensional Materials Physics and Preparing Technology, Department of Physics, Tianjin University, Tianjin 300354, China
[3]Department of Physics, Fu Jen Catholic University, Taipei 24205, Taiwan
[4]International College of Semiconductor Technology, National Yang Ming Chiao Tung University, Hsinchu 30010, Taiwan

**Abstract**

**We reanalyze the seminal work by Dolan and Osheroff (Phys. Rev. Lett. 43, 721 (1979)) which reported anomalous low-temperature conduction of high-resistivity thin-film metal strips. We argue that the observed logarithmic increase of resistance with decreasing temperature in their 3-nm-thick Au-Pd strips be ascribed to the granularity effect on electron conduction in discontinuous metal films. This reanalysis is further supported by our measurements on conducting $Pb_x(SiO_2)_{1-x}$ nanogranular films, where $x$ is the volume fraction of Pb.**

The scaling theory of localization by Abrahams *et al.*[1)] is one of the cornerstone advances in the field of quantum electron transport in disordered media of the past half century. It states that no metallic states exist in two dimensions (2D) as temperature $T \to 0$. In particular, for a 2D system in the diffusive regime ($1 \ll k_F l_e \ll \infty$, with $k_F$ being the Fermi wavenumber, and $l_e$ the electron elastic mean free path), the scaling theory predicts a small logarithmic increase of the sheet resistance ($R_\Box$) with decreasing *T*, which is known as the weak-localization (WL) effect. In practice, a ln*T* increase of $R_\Box$ can also arise from the electron-electron interaction (EEI) effect predicted by Altshuler and coworkers,[2)] which usually dominates over the WL effect in homogeneous metal films even in zero magnetic field.

The first experiment reporting a logarithmic dependence and taken to support the scaling theory

---

[a)] Corresponding author: E-mail: jjlin@nycu.edu.tw (J.J.L.)
[b)] Corresponding author: E-mail: zhiqingli@tju.edu.cn (Z.Q.L.)

was the Dolan-Osheroff's measurement on ultrathin, discontinuous Au-Pd films, which were prepared by the evaporation method.[3] In literature, it has fervently been presented that the experiment[3] worked hand in hand with theory,[1] thus fostering a good understanding of electron wave scattering in random media.[4–9] Notably, Dolan and Osheroff reported a relatively large $\ln T$ increase in $R_\square$ by about 8% as $T$ decreased in a comparatively narrow range from 2.2 to 0.32 K in a ~3-nm-thick film (denoted film C in the seminal paper[3]). This particular film had a $R_\square = 4600$ Ω (corresponding to an anomalously large resistivity $\rho \sim 1400$ μΩ cm) which was more than one order of magnitude higher than that of evaporated continuous Au-Pd films.[10-12] Dolan and Osheroff noticed the granular structure of the film and attributed the large $R_\square$ value to the tunneling resistance between metallic islands, while they and Anderson *et al.*[13] still ascribed the $\ln T$ increase to the WL effect. At the time, the Coulomb effect on electronic conduction in granular metals in the metallic regime was yet to be explored. In this work we aim to clarify the situation and point out that the low-$T$ resistance increase reported in Ref. 3 is primarily due to a many-body effect in inhomogeneous granular metals[14-16] rather than WL nor Altshuler-Aronov EEI effect.

As far as inhomogeneous granular effect is concerned, Beloborodov and coworkers[14-16] have recently studied the Coulomb interaction effect on the conductance of granular metal films above the metal-insulator transition, i.e., $g_T = G_T / (2e^2 / \hbar) > g_T^c$, where $(g_T)$ $G_T$ is the (dimensionless) tunneling conductance between neighboring grains, *e* is the electronic charge, and $2\pi\hbar$ is the Planck constant. Here the critical conductance is defined by $g_T^c = (1/2\pi d)\ln(E_c/\delta)$, where *d* is the array dimensionality of the granular metal film, $E_c$ is the charging energy, and $\delta$ is the average level spacing in the constituent metal grains. The theory of Beloborodov and coworkers predicts that in the temperature range $g_T \delta / k_B < T < E_c / k_B$ (which is pertinent to film C, where $k_B$ is the Boltzmann constant), the change in conductivity is logarithmic in $T$ for all array dimensions (*d* = 1, 2 and 3), and given by

$$\Delta\sigma(T) = \sigma(T) - \sigma(T_0) = \frac{\sigma_0}{2\pi d g_T} \ln\left(\frac{T}{T_0}\right), \qquad (1)$$

where $T_0$ is a reference temperature, and $\sigma_0$ is the conductivity without the Coulomb interaction. Note that this $\ln T$ dependence is the consequence of and specific to the granularity effect. This robust logarithmic behavior has not been addressed in Refs. 1 and 2 which focuses on homogeneous disordered systems.

We have carefully reanalyzed and found that the low-$T$ sheet conductance of film C in Ref. 3 can be well described by Eq. (1) with a single fitted parameter $g_T \simeq 2.3$, see Fig. 1. Assuming the Au-Pd grains in film C have a disc shape with ~3 nm height and ~30 nm diameter, we estimate the characteristic temperatures to be $T^* \equiv g_T \delta / k_B \sim 0.1$ K and $E_c / k_B \sim 100$ K. Thus, Eq. (1) can be applied to describe the data in Fig. 1. We further remark that below the characteristic temperature $T^*$, the granularity effect would play a progressively diminishing role, and the Coulomb correction to the conductance should recover that predicted by the EEI effect in Ref. 2 (see also Ref. 17). Conceptually, this occurs when coherent electron motion on scales larger than the grain size dominates the transport process. That is, the crossover is expected to take place at temperatures below $T^*$ where the thermal diffusion length $L_T^* = \sqrt{D^* \hbar / k_B T}$ becomes longer than the grain size $a$, with $D^*$ being the effective electron diffusion constant of the granular metal.[18] We estimate $D^* \sim 0.1$ cm$^2$/s in film C. For comparison, we point out that a crossover from the inhomogeneous granularity effect, Eq. (1), to the homogeneous Altshuler-Aronov EEI effect has explicitly been observed in ~9-nm-thick Ag$_x$(SnO$_2$)$_{1-x}$ granular films, where $T^* \sim 15$ K.[19]

The reasons why the WL effect can be ignored in film C are as follows. In 2D homogeneous disordered systems, the WL effect causes a sheet conductance increase given by $\Delta \sigma_\square = \alpha p (e^2 / 2\pi^2 \hbar) \ln(T/T_0)$, which is governed by an electron dephasing time $\tau_\varphi \propto T^{-p}$ with $p$ being a temperature exponent. The prefactor $\alpha = 1$ ($-1/2$) for weak (strong) spin-orbit interaction. In Au-Pd films, $p \rightarrow 0$ due to the "saturation of electron dephasing time" at low temperatures.[10,11] In the presence of granularity effect, Beloborodov and coworkers[14-16] have shown that the WL effect will be suppressed above a characteristic temperature $g_T^2 \delta / k_B$, which is ~0.25 K in film C. Thus, the Dolan-Osheroff results[3] cannot be ascribed to the homogeneous nor inhomogeneous WL effect. Besides, if the resistance increase in film C were due to WL effect, one would expect the film E' in Ref. 3, which had a $R_\square$ value about a factor 2 higher than that of film C, also reveal a ln$T$ dependence instead of the observed exponential temperature dependence (see the Table I in Ref. 3).

We have measured three 0.73-μm-thick co-sputtered granular Pb$_x$(SiO$_2$)$_{1-x}$ films, where $x$ is the volume fraction of Pb, to further test the validity of the prediction of Eq. (1). The average grain size of Pb is 5.3 nm, and the granular array is effectively three-dimensional.[20] The main panels of Figs. 2(a) and (b) show that a logarithmic dependence of conductivity is clearly observed in the two films with $x = 0.60$ and 0.57, respectively. (The lowest measurement temperature is 12 K to keep Pb in the normal state to avoid the superconducting fluctuations.) The straight lines are the theoretical predictions of Eq. (1). The fitted values are $g_T \simeq 2.0$ and 0.92, compared with $g_T^c \simeq 0.40$ calculated for these two films. These conductance decreases with decreasing $T$ cannot be ascribed to the Altshuler-Aronov EEI effect, which would take place below $T^*$ ($\approx 0.34$ and $\approx 0.74$ K for the $x = 0.57$ and 0.60 films, respectively). The inset of Fig. 2(a) shows the $T$ dependence of normalized conductivities for three films with $x = 0.51$, 0.57 and 0.60, as indicated. While the two films with $x = 0.57$ and 0.60 fall close to the metal-insulator transition, the $T$-dependent behavior of the film with $x = 0.51$ exhibits an insulating feature, suggesting that Pb islands are well separated and thus granular hopping conduction[14,21] has to occur at low temperatures. This conjecture is confirmed by the plot shown in the inset of Fig. 2(b), where the sample resistivity reveals a $\rho \propto \exp\left(\sqrt{\tilde{T}_0/T}\right)$ dependence below about 40 K, as expected.

In summary, we reanalyze and suggest that the $\ln T$ increase of $R_\square$ in Dolan-Osheroff's discontinuous Au-Pd films originate from the Coulomb interaction effect in granular metals rather than being a manifestation of the scaling theory of localization in the weakly disordered regime. Our measurements on $Pb_x(SiO_2)_{1-x}$ films further support the theoretical prediction of granularity effect on electron conduction in discontinuous metal films.

**Acknowledgment**

We thank S. Kirchner, P. A. Lee and S. P. Chiu for helpful discussion. This work was supported by National Science and Technology Council of Taiwan through grant number 110-2112-M-A49-015 (J.J.L.), and by National Natural Science Foundation of China through grant number 12174282 (Z.Q.L.).

**Conflict of Interest**

The authors declare no conflict of interest.

**Data Availability**

The data that support the findings of this study are available from the corresponding author upon reasonable request.

**Figure and caption**

Figure 1

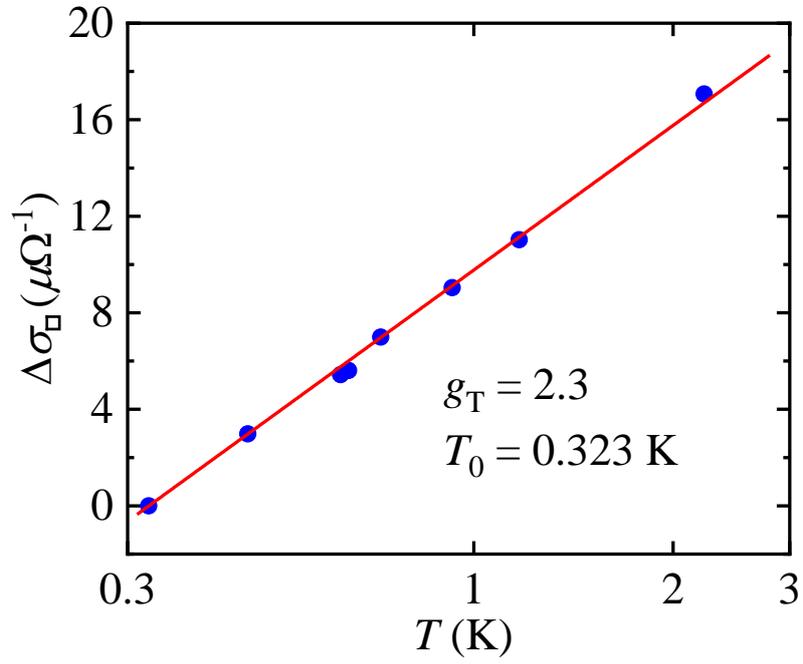

Fig. 1. Variation of sheet conductance (blue symbols) with temperature for the film C of Ref. 3. The straight line is the theoretical prediction of Eq. (1) for granular metals in the metallic regime.

Figure 2

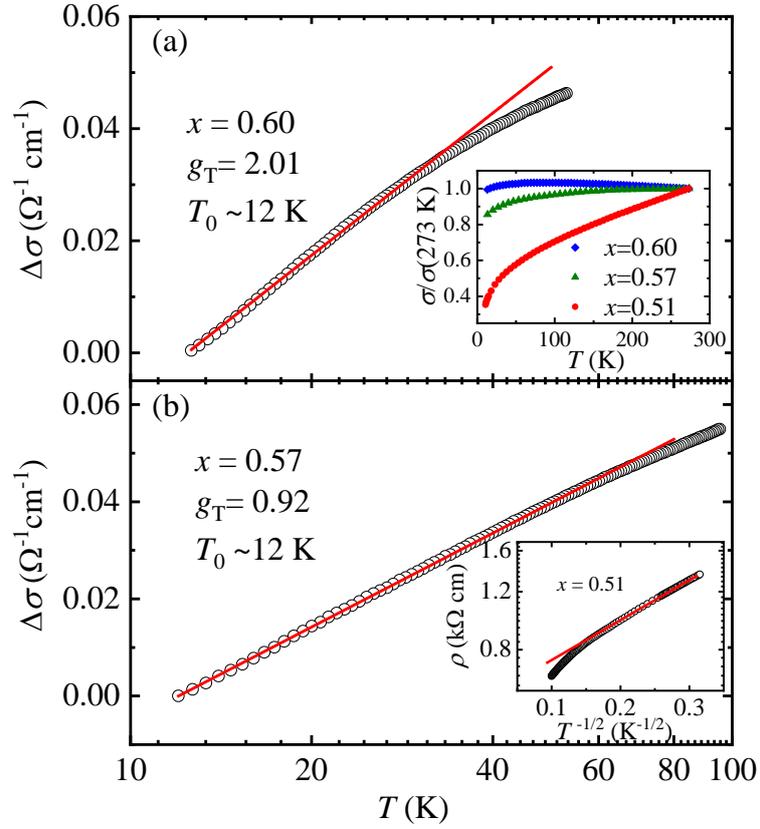

Fig. 2. Conductivity change as a function of temperature for three $Pb_x(SiO_2)_{1-x}$ granular films, with Pb volume fraction $x$ as indicated. The red straight lines are the theoretical predictions of Eq. (1). Inset of (a) shows the normalized conductivities as a function of temperature for three films. The conductivities of the $x$ = 0.60, 0.57, and 0.51 films at 273 K are 1.34, 0.48, and $2.13 \times 10^{-3}$ $\Omega^{-1}$ $cm^{-1}$, respectively. Inset of (b) shows the stretched-exponential temperature behavior of the insulating film ($x$ = 0.51), indicating granular hopping conduction.